\documentclass[%
 aip,
 amsmath,amssymb,
 reprint,%
]{revtex4-1}

\usepackage{graphicx}
\usepackage{dcolumn}
\usepackage{bm}

\usepackage[utf8]{inputenc}
\usepackage[T1]{fontenc}
\usepackage{mathptmx}
\usepackage{etoolbox}

\usepackage{soul}

\usepackage{tikz}
\usetikzlibrary{shapes.geometric, arrows, calc, positioning, decorations.pathmorphing}
\tikzstyle{lattice} = [rectangle, rounded corners, minimum width=3cm, minimum height=1cm,text centered, draw=black]
\tikzstyle{light} = [rectangle, minimum width=3cm, minimum height=1cm, text centered, text width=3cm, draw=black, fill=red!30]
\tikzstyle{machine} = [circle, minimum width=1cm, minimum height=1cm, text centered, text width=1cm, draw=black, fill=gray!30]
\tikzstyle{layer} = [rectangle, minimum width=4.5cm,minimum height=0.5cm, text width=4.2cm, thick]
\tikzstyle{arrow} = [thick,->,>=stealth]

\makeatletter
\def\@email#1#2{%
 \endgroup
 \patchcmd{\titleblock@produce}
  {\frontmatter@RRAPformat}
  {\frontmatter@RRAPformat{\produce@RRAP{*#1\href{mailto:#2}{#2}}}\frontmatter@RRAPformat}
  {}{}
}%
\makeatother
\begin{document}


\title[Overcoming Non-Radiative Losses with AlGaAs PIN Junctions for Near-Field Thermophotonic Energy Harvesting]{Overcoming Non-Radiative Losses with AlGaAs PIN Junctions for Near-Field Thermophotonic Energy Harvesting}
\author{J. Legendre}
  \email{julien.legendre@insa-lyon.fr}
\author{P.-O. Chapuis}%
\affiliation{ 
Univ Lyon, CNRS, INSA-Lyon, Université Claude Bernard Lyon 1, CETHIL UMR5008,\\ F-69621, Villeurbanne, France}

\date{\today}

\begin{abstract}
In a thermophotonic device used in an energy-harvesting configuration, a hot light-emitting diode (LED) is coupled to a photovoltaic (PV) cell by means of electroluminescent radiation in order to produce electrical power. 
Using fluctuational electrodynamics and the drift-diffusion equations, we optimise a device made of an AlGaAs PIN LED and a GaAs PIN PV cell with matched bandgaps. We find that the LED can work as an efficient heat pump only in the near field, where radiative heat transfer is increased by wave tunnelling. A key reason is that non-radiative recombination rates are reduced compared to radiative ones  in this regime. At 10 nm gap distance and for $100\text{ cm.s}^{-1}$ effective surface recombination velocity, the power output can reach $2.2 \text{ W.cm}^{-2}$ for a 600 K LED, which highlights the potential for low-grade energy harvesting.
\end{abstract}

\maketitle

Thermophotovoltaics (TPV) is a kind of solid-state heat engine (along with thermoelectrics\cite{Mao2018} and thermionics\cite{Campbell2021} for instance) where a hot emitter radiates towards a photovoltaic (PV) cell, which then converts radiation into electricity. While TPV performs well at high temperature\cite{LaPotin2022}, its capabilities drop as the hot source temperature goes below 1000 K\cite{Green2016}. To alleviate this constraint, near-field enhancements for TPV (NF-TPV) have been extensively studied\cite{Inoue2021, Lucchesi2021,Mittapally2021} (see the review by Song et al.\cite{Song2022a}), and should allow to obtain good performance down to 700 or 800 K\cite{OkanimbaTedah2019}. The decrease of the gap distance between the emitter and the PV cell opens indeed new channels for the transport of photons between the two bodies, evanescent waves being now able to participate to the heat transfer\cite{Volokitin2007}.

Another way to increase the power produced by the device is to replace the passive emitter by a light-emitting diode (LED), which is physically similar to a PV cell but used in an opposite fashion (i.e., it emits light instead of producing electricity). In an LED, the electricity consumption allows to enhance the photon emission above the bandgap: this is electroluminescence. When the energy conversion is efficient enough, the LED wall-plug efficiency (WPE), defined as the ratio between the electroluminescent (EL) radiation exchanged and the electrical power fed to the device, can actually exceed unity\cite{Santhanam2012,Radevici2019,Sadi2020}. Thermodynamically, this can happen thanks to the low entropy flux associated with EL radiation compared to thermal radiation. The LED works then as a heat pump between its phonon bath (the cold body), and the EL-enhanced electromagnetic field (the hot body)\cite{Weinstein1960,Landsberg1968,Xue2017}, the WPE corresponding to the classical coefficient of performance (COP). This is the electroluminescent cooling (ELC) regime.

\begin{figure*}
    \centering
    \includegraphics{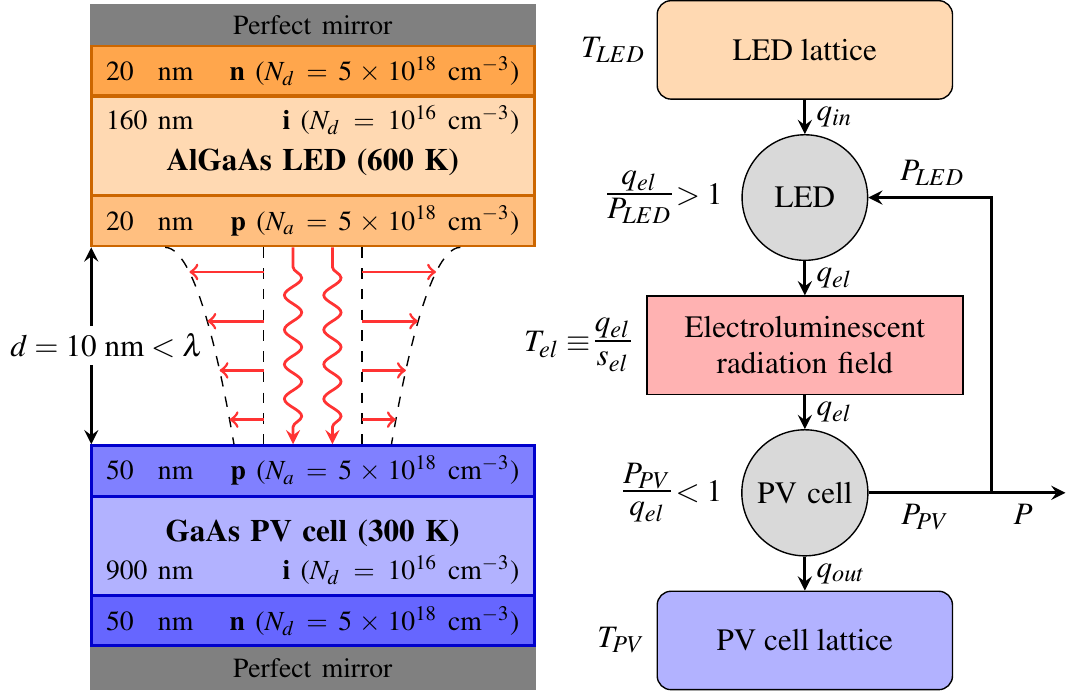}
    \caption{Physical (left) and thermodynamical (right) description of the near-field thermophotonic device considered, working in an energy-harvesting configuration.}
    \label{fig:intro}
\end{figure*}

If the conversion efficiency of the PV cell, which can be seen as a heat engine between the EL radiation field and its phonon bath, is high enough, it is then beneficial to take some electrical power from the PV cell back to the LED to enhance the above-bandgap radiation between the two bodies. The below-bandgap radiation remains unchanged, which is essential to keep the PV cell efficiency high. The combination of an LED and a PV cell in such device, as shown in Fig. (\ref{fig:intro}), is called thermophotonics (TPX)\cite{Harder2003}. It can be used both as a heat engine and a heat pump/refrigerator. NF effects allowing to reach ELC more easily, it has been combined with the EL enhancement into a NF-TPX device\cite{Zhao2018,Legendre2022,Yang2022}. Compared to TPV and NF-TPV, it can be competitive at much lower hot source temperature, down to 450-500 K, and could therefore have applications for low-grade waste heat recovery, as shown in Sec. I of the supplementary material.

In this work, we simulate the performance of a NF-TPX device composed of an AlGaAs PIN LED and a GaAs PIN PV cell (see Fig. (\ref{fig:intro})). Both the LED and the PV cell are homojunctions.While using heterostructures opens new degrees of freedom for device design, it also comes with its own issues (both in terms of fabrication and physical analysis) and we thus restrict our study to homostructures in the present work. GaAs has been chosen for the PV cell as GaAs-based structures can reach high conversion efficiency, while using AlGaAs for the LED allows to keep the bandgap of both components matched when at different temperatures, maximising the performance\cite{Zhao2018}. We place perfect mirror at the back of each component, allowing to improve radiative transfer through cavity effect and preventing radiative losses at the back surface. The impact of mirrors non-ideality on performance was studied by Zhao et al.\cite{Zhao2018}.

Modelling such devices requires to couple the resolution of the near-field photon transfer between the two components (LED and PV cell) and of the charge transport in both of them. This resolution is performed in 1D. The near-field radiative heat transfer is obtained using the fluctuational electrodynamics framework\cite{Polder1971,Rytov1989a}. The spectral photon flux density between any layers \textit{i} and \textit{j} is then expressed as:
\begin{equation}\label{eq:specPhoton}
    \gamma_{ij}(\omega)=\Delta n^0_{ij}(\omega)\mathcal{F}_{ij}(\omega).
\end{equation}

The total photon and heat flux densities are:
\begin{subequations}\label{eq:totPhoton}
\begin{gather}
    \gamma_{ij}=\int_{0}^{+\infty} \gamma_{ij}(\omega)d\omega,\\
    q_{ij}=\int_{0}^{+\infty} \hbar\omega\gamma_{ij}(\omega)d\omega.
\end{gather}
\end{subequations}

In Eq. (\ref{eq:specPhoton}), $\Delta n^0_{ij}(\omega)$ corresponds to the difference of the modified Bose-Einstein distribution functions between the two layers, expressed as
\begin{equation}\label{eq:BoseEinstein}
    n_i^0(\omega,\mu_i,T_i)=\frac{1}{\exp{\left(\frac{\hbar\omega-\mu_i}{k_bT_i}\right)}-1},
\end{equation}

where $\mu$ is the photon chemical potential, which accounts for electroluminescent enhancement. It is set to 0 for $\omega<\omega_g$ (no electroluminescence below the bandgap), and can be approximated by $\mu\approx eU$ above the bandgap. The validity of such approximation for NF-TPV applications was discussed by Callahan et al.\cite{Callahan2021}. Here, both components are assumed to be thermal reservoirs, i.e. they remain at constant temperatures. Thus Bose-Einstein distributions are uniform throughout each component, and only above-bandgap photons impact the electrical power generated or consumed.

The second factor in Eq. (\ref{eq:specPhoton}), $\mathcal{F}_{ij}(\omega)$, corresponds to the photon transmission function, and is obtained through the calculation of the electromagnetic transmission coefficient $\mathcal{T}$ following Francoeur et al.\cite{Francoeur2009}:
\begin{equation}\label{eq:transferFunc}
    \mathcal{F}_{ij}(\omega)=\frac{1}{4\pi^2}\int_0^{+\infty}\mathcal{T}(\omega,k_{||})k_{||}dk_{||}.
\end{equation}

The use of such formalism allows to account for frustrated modes (i.e., propagative modes in the components which are tunnelling through the vacuum gap) and surface modes (e.g., surface plasmon polaritons and surface phonon polaritons, which are propagating along interfaces), which enhance the radiative heat transfer in the near field.

The photon flux density being known, the current-voltage (IV) characteristic of each component should be computed. To do so, Poisson (Eq. (\ref{eq:Poisson})), drift-diffusion (Eq. (\ref{eq:Drift-Diffusion})) and continuity (Eq. (\ref{eq:Continuity})) differential equations are solved in the LED and in the PV cell:
\begin{equation}\label{eq:Poisson}
    \frac{dE}{dz}(z)=-\frac{e}{\varepsilon_s}
        \left(n(z)-p(z)+N_a(z)-N_d(z)\right),
\end{equation}
\vspace{-4mm}
\begin{subequations}\label{eq:Drift-Diffusion}
\begin{gather}
    J_n(z)=en(z)\mu_nE(z)+eD_n\frac{dn}{dz}(z),\\
    J_p(z)=ep(z)\mu_pE(z)-eD_p\frac{dp}{dz}(z),
\end{gather}
\end{subequations}
\vspace{-4mm}
\begin{subequations}\label{eq:Continuity}
\begin{gather}
    \frac{dJ_n}{dz}(z)=-e\left(G(z)-R(z)\right),\\
    \frac{dJ_p}{dz}(z)=\hphantom{-}e\left(G(z)-R(z)\right).
\end{gather}
\end{subequations}

$E$ corresponds to the amplitude of the electric field, $\varepsilon_s$ to the static dielectric constant, $n$ (resp. $p$) to the electron (resp. hole) density, $N_d$ (resp. $N_a$) to the donor (resp. acceptor) doping level, $J$ to the current density and $\mu_n$ (resp. $\mu_p$) to the electron (resp. hole) mobility, and should not be confused with the chemical potential. $D_n$ (resp. $D_p$) is the closely-related electron (resp. hole) diffusion coefficient, obtained using Einstein's relation. $G$ (resp. $R$) is the electron-hole pair (EHP) generation (resp. recombination) rate. To account for near-field effects, we assume that the radiative recombination rate equals the above-bandgap (subscript a) emission rate towards the other component, which gives, with Shockley-Read-Hall (SRH) and Auger non-radiative processes:
\begin{subequations}\label{eq:GR}
\begin{gather}
    G(z)-R(z)=-\frac{d\gamma_{a,net}}{dz}(z)-R_{SRH}(z)-R_{Auger}(z),\label{eq:GRgeneral}\\
    R_{SRH}(z)=\frac{n(z)p(z)-n_i^2}{\tau_p(n(z)+n_i)+\tau_n(p(z)+n_i)},\label{eq:SRH}\\
    R_{Auger}(z)=(C_nn(z)+C_pp(z))(n(z)p(z)-n_i^2).\label{eq:Auger}
\end{gather}
\end{subequations}

This neglects the effect of photon recycling on radiative generation and recombination rates. Since $T$ and $\mu$ are considered constant in a component, the net internal photon flux (and thus the radiative $G-R$ rate related to photon recycling) should be null.

In a previous paper\cite{Legendre2022}, we solved a simplified version of this system of coupled and nonlinear equations, with an assumption that restricts the doping levels to large values and cannot handle PIN junctions. Here, we choose to fully solve it, following Gummel's iterative method\cite{Gummel1964} and its application to TPV by Blandre et al.\cite{Blandre2017} (see Sec. III of supplementary material for more information on the charge carrier transport model). At boundaries ($z=z_b$), we assume that charge neutrality holds ($dE/dz(z_{b})=0$), that majority carriers are at equilibrium (e.g., $n_{n}(z_b)=n_n^0(z_b)$) and that minority carriers recombine following a mechanism similar to SRH, e.g. for electrons:
\begin{equation}\label{eq:surfRec}
    J_n(z_b)=e\frac{S_nS_p(n(z_b)p(z_b)-n_i^2)}{S_n(n(z_b)+n_i)+S_p(p(z_b)+n_i)},
\end{equation}

where $S_n$, $S_p$ are the electron and hole effective surface recombination velocities, and are taken equal to 100 $\text{cm.s}^{-1}$. A similar expression is used for minority holes. The main material properties given as inputs of the simulations can be found in Table \ref{tab:matprop}. Since the variation of non-radiative recombination coefficients with temperature and alloy fraction are currently unknown, they are assumed equal to those of GaAs at 300 K. We only account for alloy-fraction variation in the effective masses.

\begin{table}
\caption{Material properties}\label{tab:matprop}
\begin{ruledtabular}
\begin{tabular}{llll}
Parameter & Symbol & Value & Ref.\\
\hline
Dielectric function & $\varepsilon$ & & \\
\textit{- Reststrahlen} & & & \cite{Adachi1994,Losego2009}\\
\textit{- Interband} & & & \cite{Gonzalez-Cuevas2007}\\
Bandgap energy & $E_g$ & 1.426 eV & \cite{Gonzalez-Cuevas2007}\\
\textit{e}/\textit{h} rel. eff. mass & $m_{n/p}^*$ & & \cite{Levinshtein1999}\\
\textit{e}/\textit{h} mobility & $\mu_{n/p}$ & &\cite{Sotoodeh2000}\\
SRH recomb. coeff. & $\tau_{n/p}$ & $3.3\times 10^{-6} \text{ s}$ & \cite{Sadi2019a}\\
Auger recomb. coeff. & $C_{n/p}$ & $10^{-30} \text{ cm}^6\text{s}^{-1}$ & \cite{Sadi2019a}\\
\end{tabular}
\end{ruledtabular}
\end{table}

\begin{figure}
    \centering
    \includegraphics{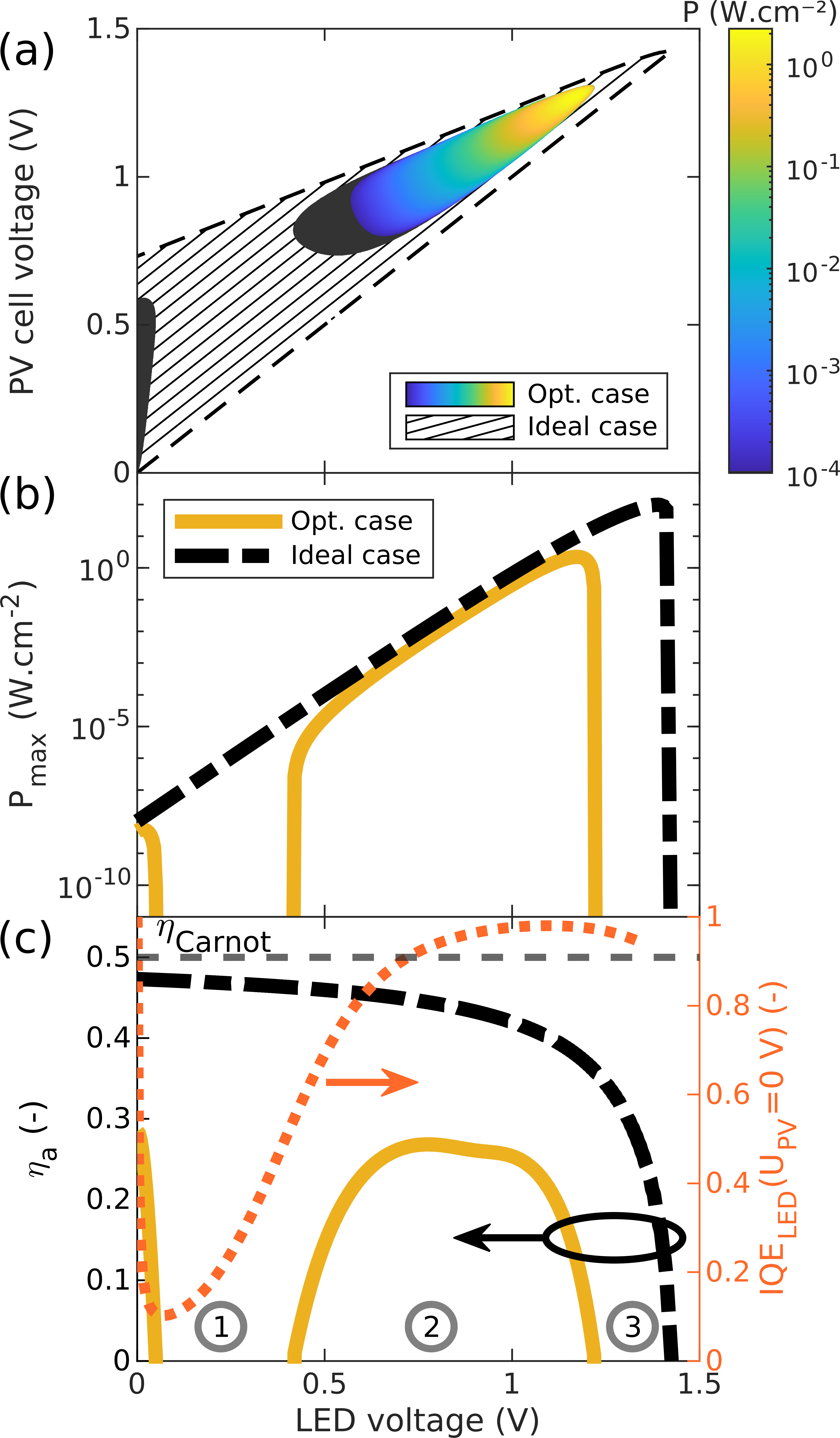}
    \caption{Capabilities of the device considered. The complete power characteristic and a simplified version are respectively given in (a) and (b). (c) is the simplified efficiency characteristic. The optimised case corresponds to the device shown in Fig. (\ref{fig:intro}), while the ideal case is obtained using a 0D charge transport model and semi-infinite components.}
    \label{fig:perf}
\end{figure}

The device considered is shown in Fig. (\ref{fig:intro}), and the thickness and doping level of each layer was obtained through an optimisation process performed at $d=10$ nm and $S=100\text{ cm.s}^{-1}$. The optimised device performance is represented in Fig. (\ref{fig:perf}) (in colour). It is compared to the results obtained with an ideal 0D charge carrier transport model (in black), where non-radiative recombinations (both in bulk and at interfaces) are neglected and both components are semi-infinite. Such model has already been extensively used for TPX \cite{Harder2003,Zhao2018,Legendre2022}. In Fig. (\ref{fig:perf}a), the full power-voltage TPX characteristic is shown, which can be simplified into Fig. (\ref{fig:perf}b) by taking for each LED voltage the maximum power output reached for the given PV cell voltage range. In Fig. (\ref{fig:perf}c), a similar approach is used to obtain the variation of the maximum above-bandgap efficiency with the LED voltage, in which below-bandgap heat transfer is not considered to emphasise the heat-to-electricity conversion. It is expressed as $\eta_a=P/q_{in,a}=P/(q_{el,a}-P_{LED})$ (see Fig. (\ref{fig:intro}) for definitions of the quantities), where $\eta_a$ is, in fact, really close to the efficiency of the maximum power points found in Fig. (\ref{fig:perf}b), with 1 to 2\% absolute difference.

\begin{figure*}[ht!]
    \centering
    \includegraphics{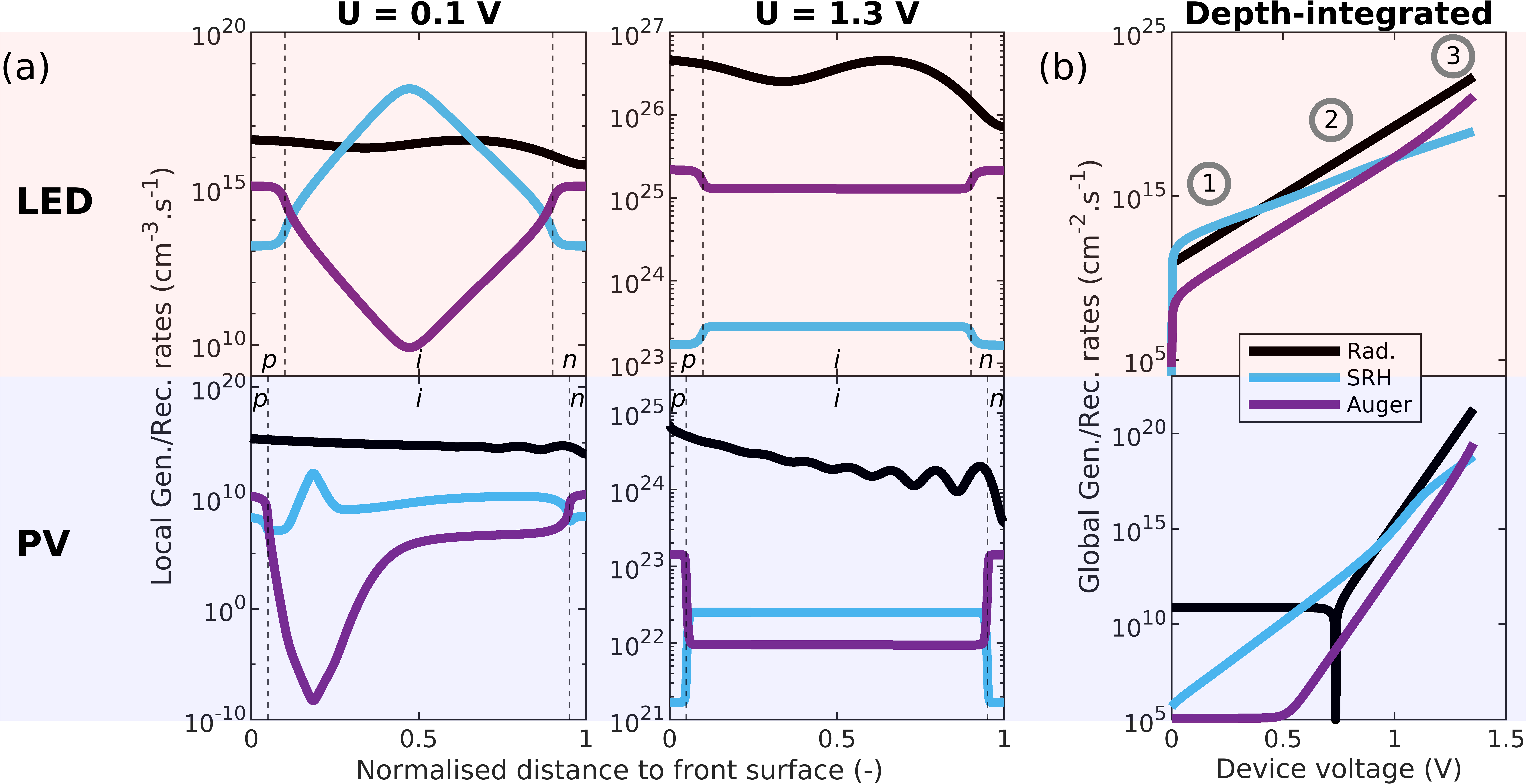}
    \caption{Comparison of SRH and Auger recombination rates with the generation-recombination radiative term, performed for the LED and the PV cell while the opposite component is unbiased. The study is done (a) as a function of the position for two different voltages, (b) as a function of the voltage for the rates integrated over the depth.}
    \label{fig:indepth}
\end{figure*}

This device is capable of producing electrical power at a rate of $2.2\text{ W.cm}^{-2}$ with an above-bandgap efficiency of around 13\%, and can deliver decent amount of power (up to $0.18\text{ W.cm}^{-2}$) with an efficiency exceeding 25\%. This maximum power output is similar to the results obtained in other articles with a 0D charge transport model where surface recombinations are neglected (but accounting for bulk non-radiative losses), with two different set of materials (AlGaAs-GaAs\cite{Zhao2018}, CdTe-InP\cite{Yang2022}). The large power output reached here also overcomes the performance obtained in our previous paper using a simplified 1D charge transport model without surface recombinations\cite{Legendre2022}: using optimised PIN junctions allows to reach similar or higher capabilities, in more realistic conditions.

The variations of the power output with voltages obtained are different to previous TPX results. Usually, the power output increases exponentially with the LED voltage, as for the ideal case in Fig. (\ref{fig:perf}b). The power production region is delimited by a conical shape (as in Fig. (\ref{fig:perf}a)), whose size depends on the significance of the losses taken into account.\cite{Legendre2022}. In the considered case however, we see large non-monotonic variations in the low-voltage region, in which the device is not capable of producing power.
The study of the LED bulk Internal Quantum Efficiency (IQE), defined as the fraction of radiative recombinations over total bulk recombinations and shown in Fig. (\ref{fig:perf}c), provides a hint about the underlying physics. Indeed, IQE shows similar trends compared to the power output, and forcing it to 1 gives back the usual exponential increase. This means that the performance deterioration at low LED voltage is caused by high non-radiative recombination rates.

To analyse this in details, we examine the dependence of SRH and Auger recombinations rates with position and voltage. These results can be found in Fig. (\ref{fig:indepth}), and are compared with the net radiative generation-recombination rate (which corresponds to the first term on the right of Eq. (\ref{eq:GRgeneral})). This study is performed for the LED and the PV cell at two different voltages in Fig. (\ref{fig:indepth}a), while keeping the opposite component unbiased. In the low-bias case, SRH recombination rates are maximum in the intrinsic layer and minimum in the doped layers, while the opposite is observed for Auger recombinations. This is still true under high bias, however their profile is much more flattened in the i-region, which is due to lower variations in the charge carrier concentrations (see Fig. (S4) in supplementary material, Sec. III). More importantly, the dependence of SRH and Auger recombination rates with voltage are different, Auger recombinations being for instance dominant in the LED i-region at high voltage while negligible at low voltage. This is summarised in Fig. (\ref{fig:indepth}b), where the variation with voltage of the quantities integrated over the thickness is shown in absolute value. While the radiative and Auger contributions have similar variations up to high voltage, the SRH one increases with a smaller rate, which is well-known for semiconductor devices (visible for instance in two-diode models\cite{Wurfel2016}). In the PV cell, recombinations only have an impact after reaching a certain voltage, since radiative generation exceeds them under low bias (corresponding to the region where the radiative and Auger terms are flat in Fig. (\ref{fig:indepth}b)). However, their effect is significant at any voltage in the LED in which radiative generation is generally negligible - the component works close to dark conditions. Therefore, three main regions can be delimited for the LED, as shown in Fig. (\ref{fig:perf}c) and (\ref{fig:indepth}b). In the first, because SRH lifetimes are not high enough, SRH processes dominate recombinations at low voltage and make the IQE really low. In the second, the highest IQE is reached due to lower increase rate of SRH and moderate Auger recombinations. In the third the latter starts to be dominant, causing a decrease of IQE. The issue for TPX devices is that for small LED IQE, the LED is too inefficient to counterbalance the losses in the PV cell. Then, the device cannot produce power anymore: this is what happens between 0.1 and 0.4 V. Such feature was not observed in previous papers. For the ones using 0D models\cite{Zhao2018,Yang2022}, SRH recombination lifetimes considered were high, Yang et al. even stating that SRH recombinations were negligible in their case. In our previous paper\cite{Legendre2022}, the high doping levels limited the SRH recombination rate.

\begin{figure}
    \centering
    \includegraphics{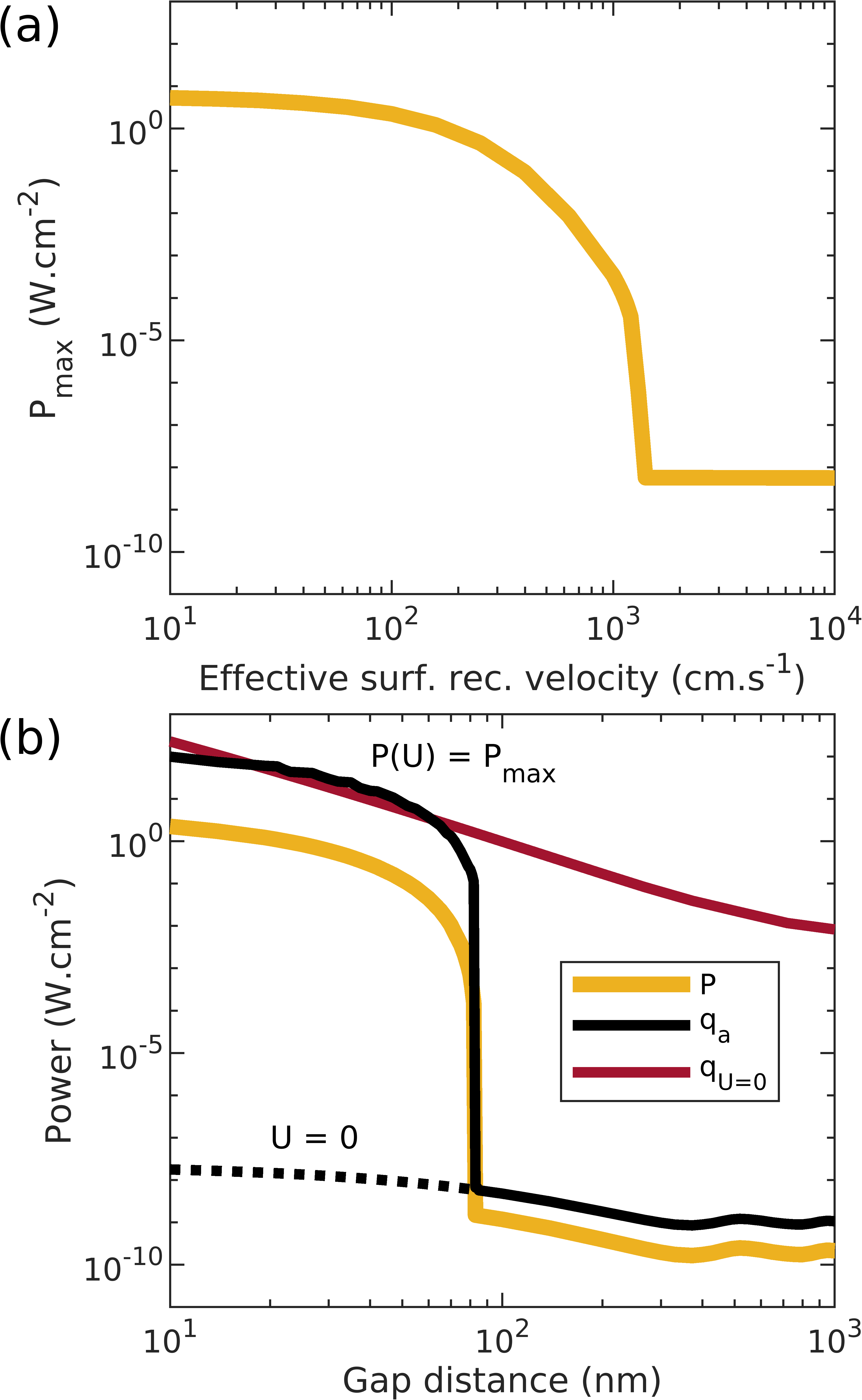}
    \caption{Impact of (a) the effective surface recombination velocity, (b) the gap distance on the device performance. In the latter, the variations of the heat flux are also analysed.}
    \label{fig:dS}
\end{figure}

The large amount of SRH recombinations at low voltage comes from the significant thickness of the i-region in both components, in which they exceed Auger recombinations as seen in Fig. (\ref{fig:indepth}a). In return, it allows to limit Auger recombinations at high voltage and thus to perform better. However, this non-monotonic behaviour will have a large impact if the working conditions are changed. This is shown in Fig. (\ref{fig:dS}), where the effective surface recombination velocity \textit{S} and the gap distance \textit{d} are changed for the given device. Similar trends can be observed for both. While previous papers reported a smooth decrease of the performance with the gap distance\cite{Zhao2018,Legendre2022}, the power output drops dramatically when using the considered device, respectively around $10^3\text{ cm.s}^{-1}$ and 100 nm. Indeed, increasing one of these quantities decreases the LED WPE. This is due either to a quicker decrease of the radiative term (directly related to $q(d)$) compared to the non-radiative ones (indirectly related to $q(d)$ via the carrier concentrations) or to an increase of interface losses. As explained previously, this decreases the region in which the TPX device can deliver power, up to the point where it is no longer beneficial to bias the LED. Then the optimal case is close to TPV (almost no LED bias), as commented in Sec. II of supplementary material. For the device considered, near-field effects are thus necessary to make the thermodynamic machine work, and cannot be interpreted as a bonus to add on top of TPX performance (as it is the case for TPV).

We now study the impact of the gap distance on the total radiated heat flux. While for the gap distances considered, mainly frustrated modes intervene for interband transitions, surface modes are important below the bandgap due to surface phonon and plasmon polaritons. As a consequence, without electroluminescent amplification, the total heat flux increases quicker than the above-bandgap heat flux when the gap distance decreases. In parallel, the above-bandgap heat flux jumps around 100 nm thanks to electroluminescence. Because of these two phenomena, the fraction of radiation located above the bandgap reaches a maximum of 65\% around 35 nm, and goes down to 30\% at 10 nm gap distance. The full device efficiency can also be computed, and is of course lower compared to its above-bandgap equivalent because of below-bandgap photons. It is equal to 2.4\% at 35 nm, and to 0.92\% at 10 nm. The complete variations of these quantities with the gap distance are shown in Sec. II of supplementary material (see Fig. S3). While these values are low, it should be possible to partially suppress the below-bandgap resonances, for instance with the use of Pt thin films\cite{Yang2022}. This would help to come closer to the above-bandgap efficiency, and limit the heat extraction issue.

In summary, the performance of an AlGaAs-GaAs NF-TPX device was simulated using fluctuational electrodynamics and an iterative resolution of Poisson, continuity and drift-diffusion equations in 1D. Compared to 0D or simplified 1D models, the study of more complex structures becomes achievable, in particular when made of weakly-doped or intrinsic layers. The realistic device considered reaches a power production of $2.2\text{ W.cm}^{-2}$ with an above-bandgap efficiency of 13\% at a 10 nm gap distance. The use of PIN homojunctions allows to increase the device performance through the limitation of Auger recombination losses and shines a light on the drastic impact of SRH recombinations on TPX characteristic at low voltage. In order to obtain even better capabilities, heterojunctions are attractive due to their high carrier selectivity and the possibility to suppress SRH recombinations through fine control of the doping profile\cite{Santhanam2020}.

\vspace{-5mm}\section*{Supplementary material}\vspace{-5mm}
See supplementary material for a comparison of NF-TPX with other technologies, a more detailed analysis of the gap distance influence, and a description of the charge carrier transport model used.

\vspace{-5mm}\section*{Acknowledgements}\vspace{-5mm}
The project TPX-Power has received funding from the European Union’s Horizon 2020 research and innovation programme under grant agreement No. 951976.
We thank P. Kivisaari and J. van Gastel for the useful discussions.

\vspace{-5mm}\section*{Author declarations}\vspace{-5mm}
\subsection*{Conflict of interest}\vspace{-5mm}
The authors have no conflicts to disclose.

\vspace{-5mm}\section*{Data availability}\vspace{-5mm}
The data that support the findings of this study are available from the corresponding author upon reasonable request.

\bibliographystyle{apsrev4-1.bst}
\bibliography{mybib.bib}

\end{document}



\title{Supplementary Material\\\normalfont{Overcoming Non-Radiative Losses with AlGaAs PIN Junctions for Near-Field Thermophotonic Energy Harvesting}} 



\author{Julien Legendre}
\author{Pierre-Olivier Chapuis}
\affiliation{Univ Lyon, CNRS, INSA-Lyon, Université Claude Bernard Lyon 1, CETHIL UMR5008, F-69621, Villeurbanne, France}

\maketitle 

\tableofcontents

\section{Comparison with performances of other devices}

\begin{figure}[h!]
    \centering
    \includegraphics{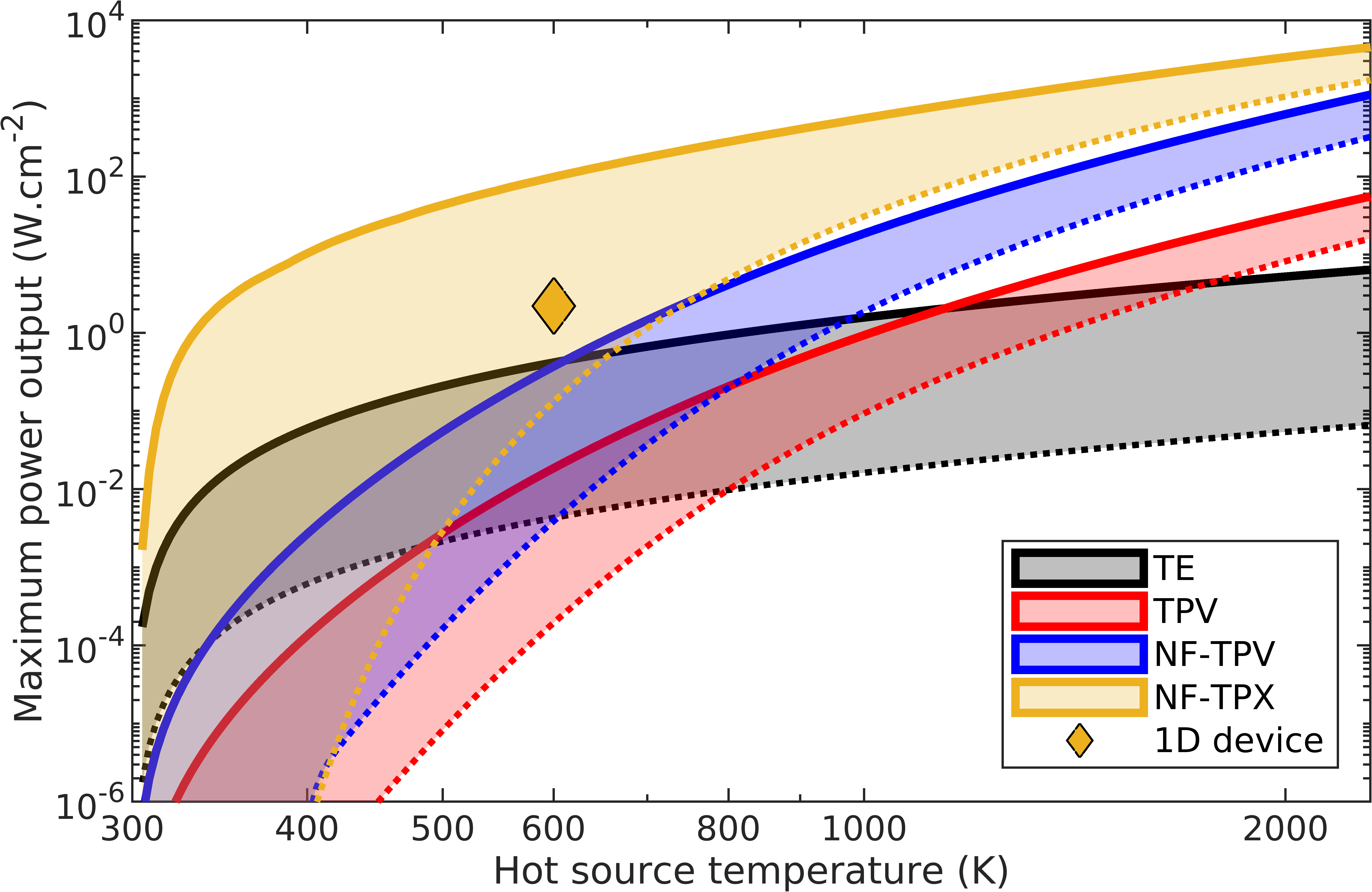}
    \caption{Variation of the power output of different devices with the hot source temperature.}
    \label{fig:Okanimba}
\end{figure}

In order to highlight the benefit of using near-field thermophotonics (NF-TPX) compared to other technologies, we show in Fig. (\ref{fig:Okanimba}) the variation of the power produced as a function of the hot source temperature for thermoelectrics (TE), thermophotovoltaics (TPV), near-field thermophotovoltaics (NF-TPV) and NF-TPX. The calculations performed are inspired by those of Okanimba Tedah et al. \cite{OkanimbaTedah2019}, with slightly different parameters. The models and values used are described in subsections (\ref{sec:subTE}) to (\ref{sec:subNFTPX}) below. One main difference is that we consider both bodies to be at constant temperature. Therefore, thermal management issues causing a decrease of performance at high hot source temperature cannot be observed here.

As mentioned in the core article, TPV can hardly compete against TE below 1000 K, while going to the near field allows to perform well down to 700 K. For NF-TPX, capabilities at low temperature are harder to predict: while the upper bound is orders of magnitude higher than for TE, the lower bound is orders of magnitude lower at very low temperatures, and is then barely better than TPV. Compared to the ideal case, the power output drops by a factor $10^7$ at 400 K, $10^4$ at 500 K, $10^3$ at 600 K for an IQE (0.9) much higher than the one used for the lower bound of TPV ($10^{-4}$). The large range of possible power outputs at low temperatures is, as explained in the article, related to the large impact of the LED IQE on the performance, which must be close to 1 in order to reach the electroluminescent cooling regime. While for TPV the device is able to produce power at any IQE, TPX devices stop working below a certain value and are thus much more sensitive to it.

Here, IQE is defined in a similar fashion to that of the article, i.e. as the fraction of recombinations being radiative. This corresponds to the usual definition of the IQE for an LED, but is quite different from its PV cell counterpart (which focuses on the incoming photon conversion efficiency). This 'LED' IQE has much lower impact on the PV cell than on the LED: this is the reason why the IQE used for TPV is so low. In addition, it should be kept in mind that while the influence of the IQE becomes lower at higher temperature, it gets harder and harder to reach high IQE at these temperatures. Moreover, thermal management issues starts to be critical both in the LED and the PV cell. We have chosen to plot NF-TPX performance up to the LED melting temperature (between 1500 K and 2000 K for AlGaAs, depending on the Al fraction), which is certainly well above realistic maximum temperature. In practice, NF-TPX devices are envisioned at low temperature.

Despite the strong effect of the LED IQE, NF-TPX can be promising for low-grade heat recovery. For an IQE of 0.9, it starts to be competitive against TE above 500 K, and any gain on the LED IQE will significantly rise the power output. The 1D case considered in the article, corresponding to the diamond in the figure, shows better performance than any other technology (even at upper bound) at 600 K, with a factor 5 enhancement compared to TE. This enhancement factor goes up to 6 and 120 for NF-TPV and TPV, respectively.

\subsection{Thermoelectrics}\label{sec:subTE}
We use the expression of the maximum power given by Apertet et al. \cite{Apertet2012}:
\begin{equation}
    P_{max}=G_{contact}\frac{Z\bar{T}}{\left(1+\sqrt{1+Z\bar{T}}\right)^2}\frac{\Delta T^2}{4\bar{T}}
\end{equation}

For the lower (resp. higher) limit, we choose $Z\bar{T}=1$ (resp 3) - corresponding to available and ambitious cases - and $G_{contact}=5 \text{ W.m}^{-2}\text{.K}^{-1}$ (resp. $G_{contact}=250 \text{ W.m}^{-2}\text{.K}^{-1}$), which are the values used by Okanimba Tedah et al. \cite{OkanimbaTedah2019}.

\subsection{Thermophotovoltaics and near-field thermophotovoltaics}
We consider that the PV cell is kept at constant temperature to be consistent with NF-TPX calculations. We assume blackbody radiation between the emitter (body 1) and the PV cell (body 2) in the case of far-field TPV. Without any non-radiative losses (upper limit) we have:
\begin{equation}\label{eq:idTPV}
    \begin{split}
        P=U\cdot J(U)&=eU\cdot \gamma_{12,a}\\
        &=eU\cdot \frac{\hbar}{4\pi^2c^2}\int_{E_g}^\infty \Delta n^0_{12}(\omega)\cdot\omega^3 d\omega\\
    \end{split}
\end{equation}

If non-radiative losses are added through a flat IQE (which is the ratio between radiative recombinations and total recombinations), this becomes:
\begin{subequations}
\begin{gather}
    P=U\cdot J(U)=eU\cdot \left(\gamma_{12,a}-\frac{1-IQE}{IQE}(\gamma_{2,a}^U-\gamma_{2,a}^0)\right)\label{eq:nidTPV}\\
    \gamma_{2,a}^U=\frac{\hbar}{4\pi^2c^2}\int_{E_g}^\infty  n^0_ 2(\omega,eU,T_2)\cdot\omega^3 d\omega
\end{gather}
\end{subequations}

In this case, we have used $E_g=0.354$ eV, corresponding to the bandgap of InAs. This low bandgap allows to obtain good capabilities at moderate emitter temperature. For the lower bound, we use IQE of $10^{-4}$, which allows to obtain a power output similar to that of Milovich et al. \cite{Milovich2020} in the far-field at 800 K.

For the near-field effects, we assume a simple enhancement factor equal to 20 compared to the far-field value. This factor corresponds to the increase of the above-bandgap heat flux at 10 nm gap distance. To compute this, we considered similar input data to those of Milovich et al. except for the InAs dielectric function \cite{Adachi1989}. Note that for a 100 nm gap distance, they obtained a $\times 30$ power enhancement factor. A more complete calculation, including charge carrier transport, would be required to have a better estimation of this factor at 10 nm.

\subsection{Near-field thermophotonics}\label{sec:subNFTPX}
A similar description to the one used for TPV is applied for NF-TPX. Both the LED (body 1) and the PV cell (body 2) are still considered semi-infinite. However, we use AlGaAs and GaAs (therefore no longer radiating at the blackbody limit), and the near-field effects are precisely computed at a distance of 10 nm - in a equivalent manner to the 1D resolution introduced in the article.

The expressions obtained are similar to Eq. (\ref{eq:idTPV}) and (\ref{eq:nidTPV}), except that they should be computed both for the LED and the PV cell. Without any non-radiative recombinations (upper limit), it gives back the '0D ideal case' used in the article. The lower limit is obtained for a flat IQE of 0.9 applied to the LED and the PV cell. As previously explained, this only has a small influence on the PV cell performance, and mainly has an effect on the LED.

\section{Zoom on the variation of the device performance with the gap distance}\label{sec:iqed}

\begin{figure}[h!]
    \centering
    \includegraphics{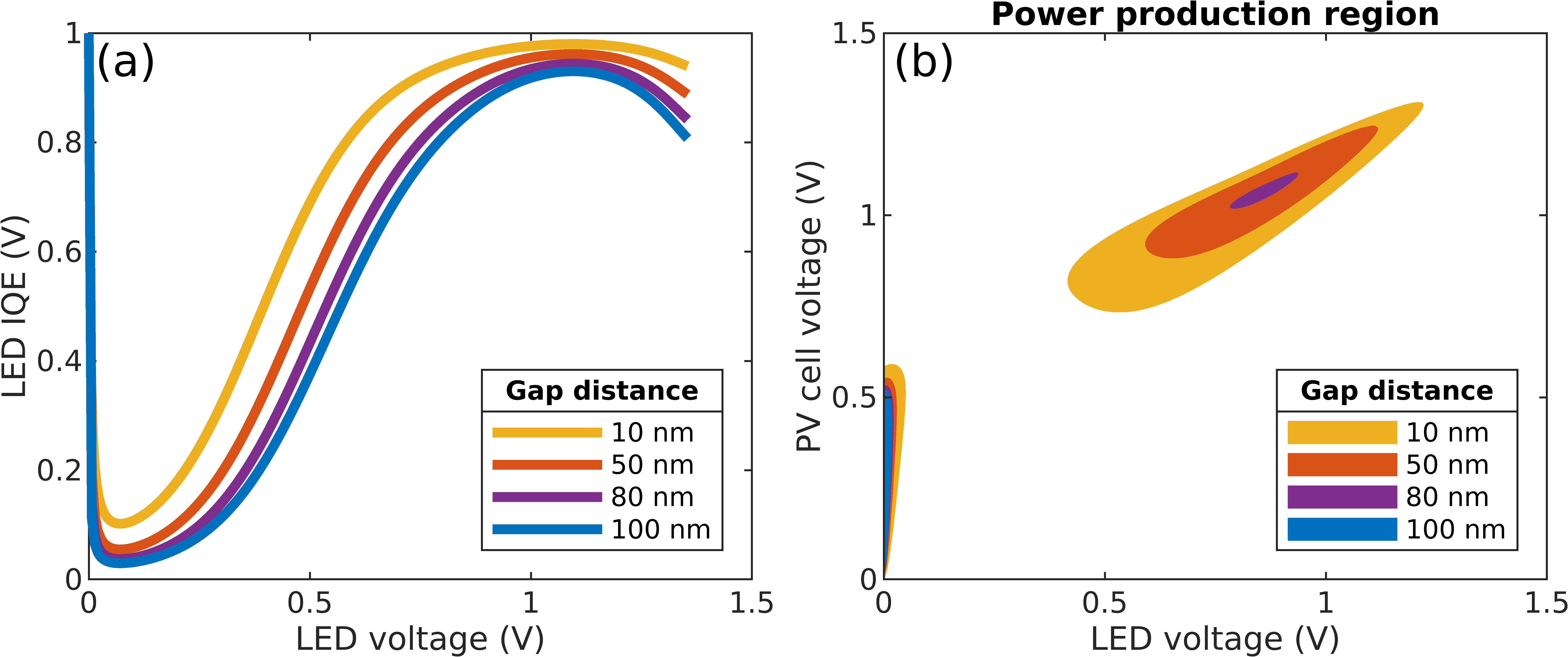}
    \caption{Variation of (a) the LED IQE (at null PV cell voltage), (b) the device full characteristic with the gap distance.}\label{fig:env_d}
\end{figure}

We aim at giving more insight on the sudden drop of power output when the gap distance is increased up to 100 nm (shown in Fig. (4) of the article). As seen in Fig. (\ref{fig:env_d}a), increasing \textit{d} lowers the LED IQE. This happens mainly because the radiative term of Eq. (8a) in the article (directly related to the gap-dependent heat flux $q(d)$) is decreasing at a higher rate than the non-radiative term (only indirectly related to $q(d)$ via the carrier concentrations). This drop is larger in the transition region between low and high IQEs, reaching 30\% in absolute value at 0.5 V.
While the decrease is lower at higher voltage (5 to 15\%), this is enough to largely reduce the power production region (PPR), as shown in Fig. (\ref{fig:env_d}b). With this figure, the brutal change of performance with distance becomes clearer: up to 80 nm, the LED IQE is still large enough for a PPR to exist under high voltage, making the variation of power output relatively smooth. However, between 80 nm and 100 nm gap distance, the LED wall-plug efficiency (WPE), defined as the ratio of the input electrical power over the radiated power, becomes too low to counterbalance the PV cell losses and the TPX device cannot produce power anymore under high LED voltage. Thus, only the close-to-TPV PPR remains (close to the y-axis), in which the power output is low.
It is interesting to notice that while the maximum LED IQE is always reached at a LED voltage close to 1.1 V, the PPR is actually centered around 0.9 V. Since the PPR is related to energy balance (rather than carriers), quantities such as the LED WPE or the PV cell efficiency are better metrics to predict its position. For instance, for a 80 nm gap distance and a 1.1 V PV cell voltage, the WPE reaches a maximum close to an LED voltage of 0.83 V. The PV cell being more efficient under higher radiative heat flux (i.e., at higher LED voltage), the optimum LED voltage is obtained as a trade-off between the LED and PV cell efficiencies and is located close to 0.9 V. 

\begin{figure}
    \centering
    \includegraphics{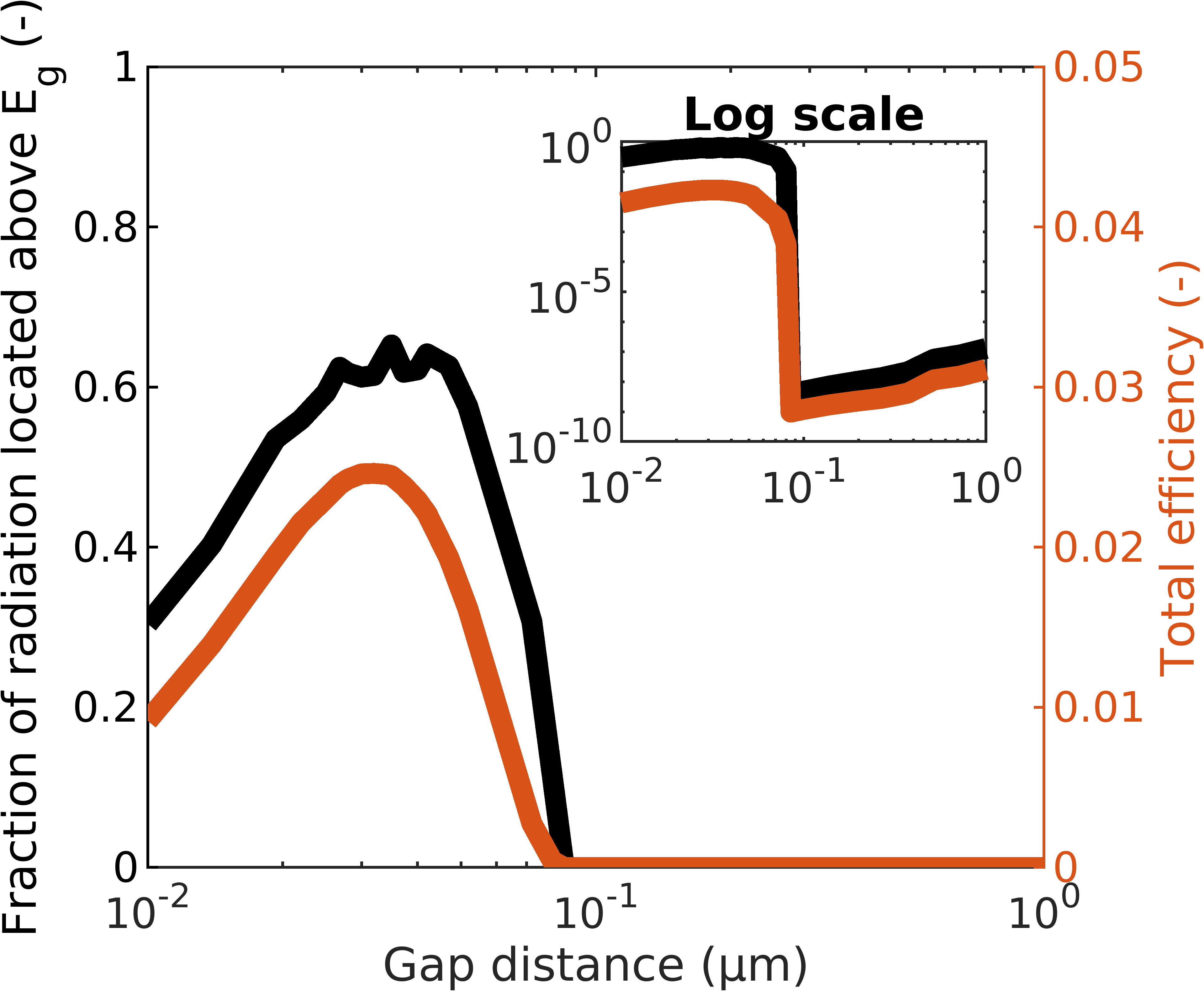}
    \caption{Variation of the fraction of radiation located above the bandgap (left axis) and of the total efficiency (right axis) as a function of the gap distance.}\label{fig:eff_d}
\end{figure}

To complete the discussion about the fraction of radiation located above the bandgap $\alpha$ discussed in the main article, we show its variation with the gap distance along with the total efficiency in Fig. (\ref{fig:eff_d}). This makes the correlation between the two quantities quite clear. A maximum is found between 30 and 40 nm, with a close-to-logarithmic decrease further away from this region. The small oscillations of $\alpha$, which can be seen around the maximum, are probably due to cavity effect. Such features can also be seen on the above-bandgap heat flux in Fig. (4a) of the main article, although less visible than in Fig. (\ref{fig:eff_d}).

\vspace{-1mm}\section{Details of the charge carrier modelling}
\subsection{Numerical model}

As stated in the core article, we iteratively solve Poisson, continuity and drift-diffusion equations. To do so, we use the so-called Slotboom variables $\Phi_n$, $\Phi_p$ \cite{Slotboom1973} and the normalised electrostatic potential $\Psi$. The carrier concentrations are then expressed as

\begin{subequations}
\begin{gather}
    n=n_i \exp{\left(\frac{eV}{k_bT}\right)} \exp{\left(-\frac{E_{Fn}}{k_bT}\right)}=n_i \exp{\left(+\Psi\right)} \Phi_n,\\
    p=n_i \exp{\left(-\frac{eV}{k_bT}\right)} \exp{\left(\frac{E_{Fp}}{k_bT}\right)}=n_i \exp{\left(-\Psi\right)} \Phi_p,
\end{gather}
\end{subequations}

where $E_{Fn}$, $E_{Fp}$ are the associated quasi-Fermi levels and $V$ is the electrostatic potential. This allows to express the system of equations as follows:

\begin{equation}\label{eq:PoissonSimp}
    \frac{d^2\Psi}{d\Bar{z}^2}(z)=\left(\exp{\left(\Psi(z)\right)}\Phi_n(z)-\exp{\left(-\Psi(z)\right)}\Phi_p(z)+N\right(z)),
\end{equation}
\vspace{-1cm}\begin{subequations}\label{eq:DriftDiffusionSimp}
    \begin{gather}
        J_{n}(z)=+e\cdot n_i D_{n}\exp{\left(+\Psi(z)\right)}\cdot\frac{d\Phi_n}{dz}(z)=a_n(\Psi(z))\frac{d\Phi_n}{dz}(z), \\
        J_{p}(z)=-e\cdot n_i D_{p}\exp{\left(-\Psi(z)\right)}\cdot\frac{d\Phi_p}{dz}(z)=a_p(\Psi(z))\frac{d\Phi_p}{dz}(z),
    \end{gather}
\end{subequations}
\vspace{-1cm}\begin{subequations}\label{eq:ContinuitySimp}
    \begin{gather}
        \frac{dJ_{n}}{dz}(z)=+e(R(z)-G(z)), \\
        \frac{dJ_{p}}{dz}(z)=-e(R(z)-G(z)).
    \end{gather}
\end{subequations}

$N$ corresponds to the doping level, and is negative for donors, while $\bar{z}$ is the position normalised by the Debye length. The general resolution method has been explained in several articles or textbooks \cite{Gummel1964,Vasileska2017}. Solving alternatively Poisson equation (to obtain $\Psi$) and continuity and drift-diffusion equations (to obtain $\Phi_n$ and $\Phi_p$) using the previous solution as input for unknown quantities, we are able to obtain self-consistent solutions. Poisson equation is solved using linearisation and an iterative process.

Regarding the resolution of continuity and drift-diffusion equations, several methods can be used \cite{Scharfetter1969,Blandre2016,Blandre2017}. In our case, we choose to solve them via integration to reduce the risk of numerical instability. To do so, we first compute the electron and hole currents using continuity equations and the value given by the recombination current at one of the interfaces, which gives for instance for electrons:

\begin{equation}
    J_n(z)=J_n(z_b)+e\int_{z_b}^{z}(R(z')-G(z'))dz'.
\end{equation}

Then, the Slotboom variables are obtained using drift-diffusion equations and the condition of equilibrium at contacts for majority carriers (which gives $\Phi$ at one of the interfaces), e.g. for electrons:

\begin{equation}
    \Phi_n(z)=\Phi_n(z_b)+\int_{z_b}^{z}\frac{J_n(z')}{a_n(\Psi(z'))}dz'.
\end{equation}

In order to help convergence, after each iteration in which we solve the whole system, the obtained solution is averaged with the previous one using specific weights. We verified that such process did not change the final solution.

\subsection{Comparison with low-injection approximation}

In order to show the interest of solving fully the system of equations rather than using approximations (as in our previous article \cite{Legendre2022}), we compare our results with those obtained with the low-injection approximation (LIA). To simplify the study, we do this comparison only on Shockley-Read-Hall (SRH) recombination rates. In the LIA, the doping level is supposed to be high enough so that the majority carrier density is constant in the layer. For instance, for n-doped layers, the SRH expression given in Eq. (8a) of the main article becomes:

\begin{equation}\label{eq:SRHlia}
    R_{SRH}(z)=\frac{(N_dp-n_i^2)}{\tau_pN_d}=\frac{\Delta p}{\tau_p},
\end{equation}

where $\Delta p$ corresponds to the difference of hole density compared to equilibrium. 

\begin{figure}
    \centering
    \includegraphics{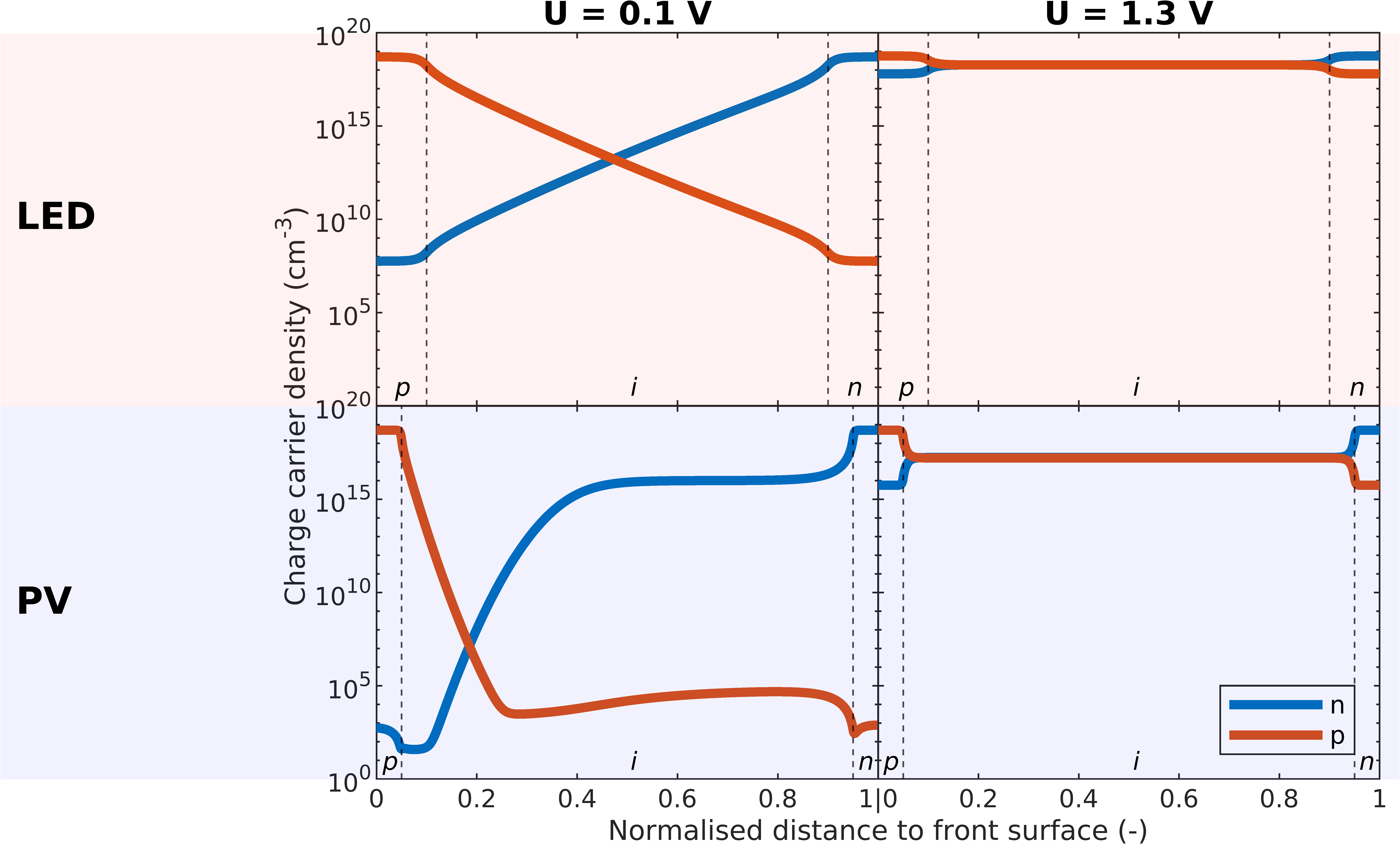}
    \caption{Charge carrier densities in the LED and in the PV cell at different voltages with the opposite component kept at 0 V.}
    \label{fig:np}
\end{figure}

\begin{figure}
    \centering
    \includegraphics{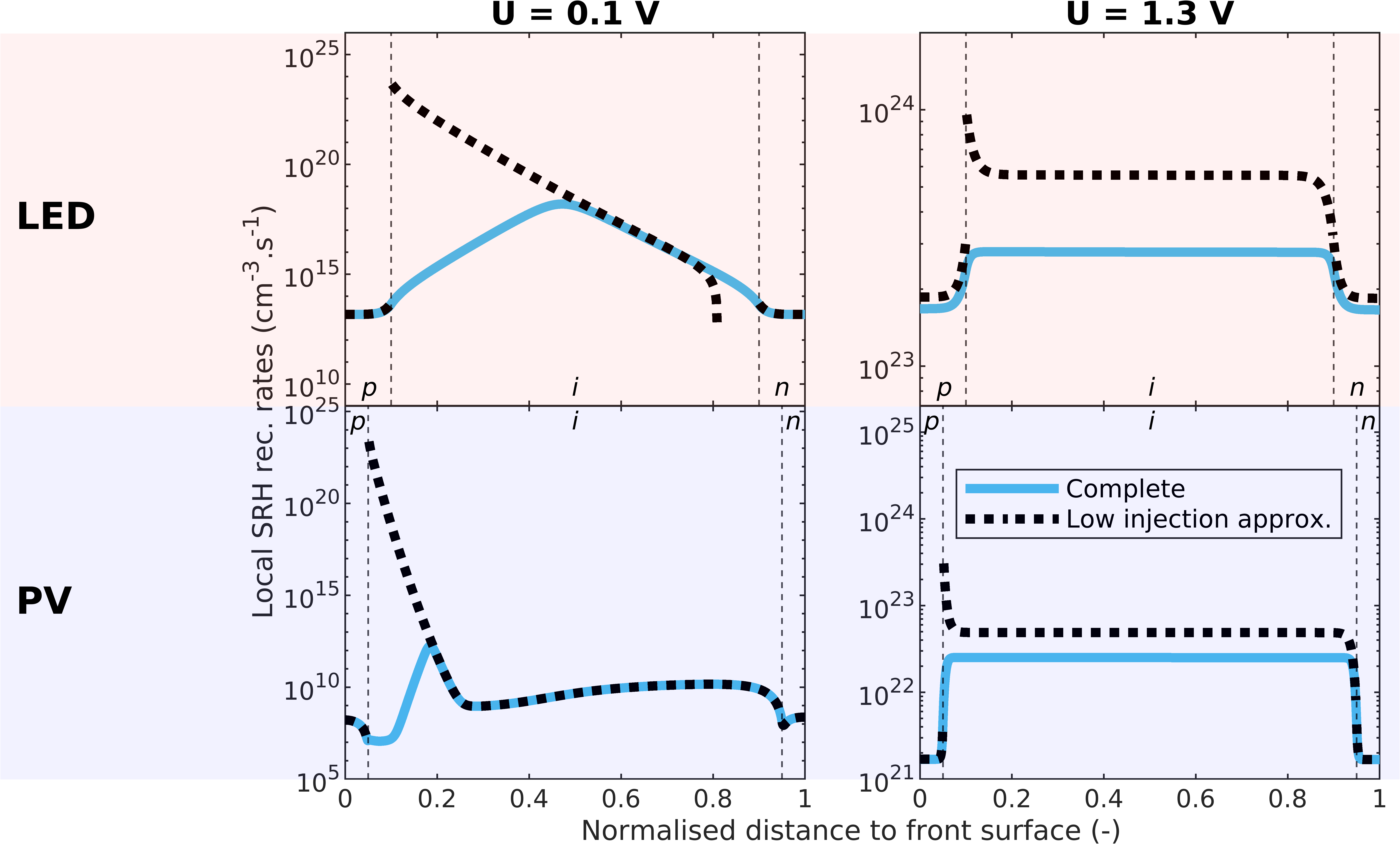}
    \caption{Comparison of SRH recombination rates obtained with the charge carrier densities shown in Fig. (\ref{fig:np}), with the complete expression or with the low-injection approximation. The study is performed for the LED and the PV cell while the opposite component is unbiased.}
    \label{fig:lia}
\end{figure}

We consider the same cases as the ones shown in Fig. (3) of the main article (i.e. with only one component being biased) using the charge carrier densities obtained after convergence of the complete model (shown in Fig. (\ref{fig:np})). The results obtained are represented in Fig. (\ref{fig:lia}). The LIA works well in the highly doped regions (at the boundaries of the junctions), even if it starts to diverge for the LED at high voltage due to large photon emission. In the intrinsic region however, it becomes erroneous and can be off by orders of magnitude compared to the complete resolution. This is due to the electron density deviating from $N_d=10^{16}\text{ cm}^{-3}$. Expressions such as Eq. (\ref{eq:SRHlia}) thus cannot be used for PIN junctions. When using the LIA, the depleted layer is often resolved analytically to prevent such issue; however, this is not possible here, since the PI and IN junctions overlap each other. For instance it can be observed that in the LED, at low voltage, the electron concentration does not stabilise to $N_d$ in the intrinsic layer (while it does in the three other cases). 

\bibliography{mybib.bib}